\documentclass[aps,twocolumn,showpacs,showkeys,floatfix,longbibliography,superscriptaddress]{revtex4-1}

\usepackage{multirow}
\usepackage{graphicx}

\usepackage{textcomp}

\usepackage{dcolumn}

\usepackage{amsmath}

\usepackage{amssymb}

\usepackage{epstopdf}

\usepackage{rotating}

\usepackage{color}

\usepackage{natbib}

\usepackage{url}

\setcitestyle{square}

\begin{document}

\title{Systemic inequities in introductory physics courses: the impacts of learning assistants}



\author{Ben Van Dusen}\affiliation{Department of Science Education, California State University Chico, Chico, CA, 95929, USA} 

\author{Jayson M. Nissen}\affiliation{Department of Science Education, California State University Chico, Chico, CA, 95929, USA}

\begin{abstract}

Creating equitable performance outcomes among students is a focus of many instructors and researchers. One focus of this effort is examining disparities in physics student performance across genders, which is a well-established problem. Another less common focus is disparities across racial and ethnic groups, which may have received less attention due to low representation rates making it difficult to identify gaps in their performance. In this investigation we examined associations between Learning Assistant (LA) supported courses and improved equity in student performance. We built Hierarchical Linear Models of student performance to investigate how performance differed by gender and by race/ethnicity and how LAs may have moderated those differences. Data for the analysis came from pre-post concept inventories in introductory mechanics courses collected through the Learning About STEM Student Outcomes (LASSO) platform. Our models show that gaps in performance across genders and races/ethnicities were similar in size and increased from pre to post instruction. LA-support is meaningfully and reliably associated with improvement in overall student performance but not with shifts in within-course performance gaps.

 \end{abstract}

\maketitle

\section{Introduction}

Disparities in student performance in science classes have been recorded throughout the United States' educational systems \citep{NationalResearchCouncil2011ExpandingCrossroads}. The PER community has a significant number of publications \citep{NationalResearchCouncil2012DisciplineEngineering} documenting the impact of course transformations on student performance overall, but has only begun to disaggregate these findings across student demographics \citep{Brewe2016EditorialPhysics}. The National Research Council (NRC) report examining the state of Discipline Based Education Research \citep[pg.~ 136-137] {NationalResearchCouncil2012DisciplineEngineering} states that while, ``DBER clearly indicates that student-centered instructional strategies can positively influence students' learning\ldots Most of the studies the committee reviewed were not designed to examine differences in terms of gender, ethnicity, socioeconomic status, or other student characteristics.'' The NRC identifies examining performance of students from underrepresented cultures as an important direction for future research.
\par In one of the few investigations of equity for both gender and race/ethnicity in physics, \citet{Brewe2010TowardPhysics} found that, while all students learned significantly more in courses with student-centered pedagogies than in courses with lecture-based instruction, gender differences increased from pre to posttest on conceptual inventories in both types of courses. However, \citet{Brewe2010TowardPhysics} also found that differences between majority and ethnic minority students did not increase in courses that used student-centered pedagogies while they did increase in courses that used lecture-based instruction.
\par 	The Learning Assistant (LA) model is one method for supporting the adoption and dissemination of student-centered instructional strategies in college science courses. An LA is an undergraduate student who are guided by course instructors and a special pedagogy course to facilitate discussions among groups of students in a variety of classroom settings that encourage student engagement and responsibility for learning. Over the last 10 years, the LA model has spread from a handful of physics and astronomy courses in 3 institutions to a range of STEM courses across 70+ institutions \citep{Otero2016LearningNetworks}. The broad dissemination of the model has led the LA-using institutions to create a support network called the LA Alliance. The LA Alliance has made it possible to measure the impact of LAs across institutional contexts.
\par 	LAs are associated with improved student performance in physics courses at multiple institutions \citep{Pollock2012ImpactsPhysics,Goertzen2011MovingTutorials}. The impacts of LAs on equity is less clear. \citet{Kost-Smith2010,Kost-Smith2012ReplicatingChallenges} found that gender differences in concept inventory scores increased from pre to posttest. Their study, however, focused on a single institution that has highly effective physics courses and may not represent the larger set of LA-using courses.
\par 	The emergence of the LA Alliance and subsequent large-scale data collection using the LASSO platform enabled researchers to collect data from large enough samples to make reliable claims about the impact of LAs on physics students from underserved backgrounds. Using the first semester of data collected using LASSO, Van Dusen et al. \citep{VanDusen2015LearningFindings} performed an exploratory analysis of the impacts of LAs across the STEM disciplines (1,645 students in 15 courses). In their analyses, Van Dusen et al. found evidence of persistent gender and racial inequities in LA-supported courses. Van Dusen et al. \citep{VanDusen2016TheOutcomes} performed a follow-up study examining the impact of LA-supported environments on first and second semester physics students from dominant and underserved backgrounds (2,868 students in 67 courses). Students from dominant backgrounds were defined as white or Asian, non-Hispanic, and male. They found that in LA-supported courses underserved students had larger shifts in their knowledge, measured as an effect size using Cohen’s \emph{d}, than dominant students. In courses without LAs, however, they found the opposite trend; dominant students had larger effect sizes. The investigation presented in this publication builds on the \citet{VanDusen2016TheOutcomes} findings by focusing on first semester physics courses, increasing the statistical power, and creating more nuanced hierarchical linear models.

\section{Purpose and Research Questions}

\par 	This study examines the role of LAs in supporting equity in college physics courses. To do this, we investigate the following research questions: (1) What gaps exist across gender and racial/ethnic student demographics in introductory physics courses? (2) How do student performance gaps in LA-supported and non-LA-supported introductory physics courses compare?

\section{Methods}

\textit{Data collection}: We accessed our large-scale, multi-institution data from the Learning About STEM Student Outcomes (LASSO) platform. The LASSO platform hosts, administers, scores, and analyzes student pre and posttest assessments online. Instructors download a report on their students' performance and have access to all of their students’ responses. Data from the courses are added to the LASSO database where they are anonymized, aggregated with similar courses, and made available to researchers with approved IRB protocols. Prior to taking the assessment on line, students are asked to complete a brief demographics questionnaire. For this study, we examined data from courses that used the Force Concept Inventory (FCI) \citep{Hestenes1992ForceInventory} or Force and Motion Conceptual Evaluation (FMCE) \citep{Thornton1998AssessingCurricula}. We did not differentiate between the FCI and FMCE in the models we present because our preliminary analysis showed that doing so did not meaningfully change the model. 

\par \textit{Data processing}: We removed assessment scores for students if they took less than 5 minutes on the assessment or completed less than 80\% of the questions. We removed entire courses if they had less than 40\% student participation on either the pre or posttest. After cleaning the data we used hierarchical multiple imputation (HMI) with the hmi and mice packages in R to address missing data. HMI is a principled method for maximizing statistical power by addressing missing data while taking into account the structure of the data. HMI also can help ameliorate selection effects from participation rates skewing toward higher performing students \citep{JariwalainpressIn-classAssessments}. HMI addresses missing data by (1) imputing each missing data point \emph{m} times to create \emph{m} complete data sets, (2) independently analyzing each data set, and (3) combining the \emph{m} results using standardized methods \citep{Drechsler2015MultipleSimplicity}. After filtering but prior to running HMI our data was  missing 15\% of the pretest scores and 30\% of the posttest scores. The analysis used 10 imputed datasets.

\par 	After cleaning and imputation, our dataset included 4,365 students from 93 courses. We used student’s self-reported demographic data to classify them using their gender (male = dominant gender; female or non-binary = underserved gender) and racial/ethnic identities (white or Asian and non-Hispanic = dominant race/ethnicity; neither white, Asian, or Hispanic = underserved race/ethnicity). In our sample, 1,662 (38.1\%) of the students are of an underserved gender and 961 (22.0\%) are of an underserved race/ethnicity. 56 (60.2\%) of the courses were LA-supported and 79 (84.9\%) of the courses used the FCI.

\par	\textit{Data analysis}: We calculated descriptive statistics to identify gaps between the average pre and posttest scores across student demographics. There were meaningful disparities in student pretest and posttest scores across student genders and races/ethnicities (Tab. 1). The gaps that students began the course with (11.9\% for underserved gender and 8.1\% for underserved race/ethnicity) are even wider by the posttest (12.3\% for underserved gender and 12.0\% for underserved race/ethnicity). These differences were within the range of gap sizes measured by \citet{Brewe2010TowardPhysics} for gender and for race.

\bgroup
\def\arraystretch{1.3}
\begin{table}
\caption{Student demographics and performance}
\begin{tabular}{p{1.4cm}lccccc}
\hline \hline
			&&&\multicolumn{3}{c}{Mean Score (\%)}&\\
\multicolumn{2}{c}{Demographics}			&N&Pre(S.D.)&&Post(S.D.)&Gain\\ \hline
\multirow{3}{*}{Gender}	&Dominant	&2,703&41.6(20.7)	&&61.9(23.7)&20.4\\
			&Underserved&1,662&29.7(17.0)	&&49.7(23.4)&20.0\\\cline{2-7}
			&Difference &-&-11.9 			&&-12.3		&-0.4\\ \hline
\multirow{3}{*}{\shortstack[l]{Race/ \\Ethnicity}}	&Dominant	&3,404&38.8(20.6)	&&59.9(24.1)&21.1\\
			&Underserved&961&30.8(17.2)		&&47.9(22.9)&17.2\\\cline{2-7}
			&Difference &-&-8.1 			&&-12.0		&-3.9\\
\hline \hline

\end{tabular}
\end{table}
\egroup




\par	To identify gaps in student performance, we used the HLM 7 software to create models that take the structure of data into account. Specifically, we developed 2-level Hierarchical Linear Models (HLM) that nest student data within course data. Our HLM models allowed us to quantify the interaction effect between a course being LA-supported and student demographic data while accounting for inherent and unknown course-level variations (e.g. the time of day of a class, student majors, and instructor backgrounds can lead to unforeseeable differences in student performance). 

\par	We developed our HLM models through a series of incremental additions of variables. In this paper we show the results from three models with postscore as the outcome variable. Model 1 is the unconditional model with no predictor variables. Model 2 includes the student (level-1) variables (gender, race/ethnicity, and student prescore). Model 3 builds on Model 2 by including the course (level-2) variables (LA-Supported and class mean prescore) and is shown below.  The level-1 equation includes a coefficient for the intercept ($\beta_{0j}$), for the underserved gender ($\beta_{1j}$), underserved race/ethnicity ($\beta_{2j}$), student prescore ($\beta_{3j}$), and for a random effects variable ($r_{0j}$). Each coefficient in level 1 has an associated level 2 equation. In the level 2 equation, the intercept is $\gamma_{i0}$, there is an associated coefficient ($\gamma_{ij}$) for each variable in the equation and $u_{ij}$ represents the random effect.

\textbf{Level-1 Equation}
\begin{eqnarray*} 
(\text{Postscore})_{ij} & = & \beta_{0j} + \beta_{1j} * (\text{Und.serv. Gender})_{ij} +\\
& &  \beta_{2j}*(\text{Und.serv. Race/Ethnicity})_{ij} +\\ 
& &  \beta_{3j}*(\text{Student Prescore})_{ij} +r_{ij}
\end{eqnarray*}

\textbf{Level 2 Equations}
\begin{eqnarray*}
\beta_{0j}& = & \gamma_{00}+\gamma_{01}*(\text{LA-Supported})_{j}) +\\ 
& & \gamma_{02}*(\text{Class Mean Prescore})_{j} +u_{0j}\\ 
\beta_{1j}&=&\gamma_{10}+\gamma_{11}*(\text{LA-Supported})_{j} + u_{1j}\\ 
\beta_{2j}&=&\gamma_{20}+\gamma_{21}*(\text{LA-Supported})_{j} + u_{2j}\\ 
\beta_{3j}&=&\gamma_{30} +u_{3j}
\end{eqnarray*}

\par LA-support is not included in the level-2 equation for student prescore because the interaction between the variables are not of interest in our analysis. For ease of interpretation, student prescore is group mean centered, class mean prescore is grand mean centered, and all other variables are uncentered. We included prescores in the model because they are strong predictors of student performance and improved the model's fit. Since prescores are not the focus of this investigation we will not discuss them in our interpretation of the models. 

\par 	We compared Model 1's level-1 (\textit{r}) and level-2 intercept variance (\textit{u\textsubscript{0j}}) (Tab. 2) to calculate the Intraclass Correlation Coefficient (ICC). In our case, the ICC identifies what percentage of the differences in student performance is attributed to student features (gender, race/ethnicity, and student prescore) versus course features (LA-support and class mean prescore). The ICC shows that course-level features explain 29\% of the variance in student performance and student-level features explain the remaining 71\%. These percentages show that course features have a substantial effect on student performance and HLM is an appropriate method of analysis. The reduction in level-1 variance (\textit{r}) from Model 1 to Model 2 (Tab. 2) shows that our student-level variables explain 27\% of the within-class variance in student performance. The reduction in level-2 intercept variance (\textit{u\textsubscript{0j}}) from Model 1 to Model 3 shows that our final model explains 57\% of the variance in mean performance across classes. The reduction in the variances associated with gender (\textit{u\textsubscript{1j}}) and race/ethnicity (\textit{u\textsubscript{2j}}) from Model 2 to Model 3 show that our final model explains 67\% of the gender gap and 27\% of the race/ethnicity gap across classes. The explained  variance shows that Model 3 has strong explanatory power. As Model 3 is our most robust model, we will focus on it in our findings section. 

\begin{table}
\caption{Hierarchical Linear Models}
\begin{tabular}{p{2.1cm}ccp{.03cm}ccp{.03cm}cc}
\hline \hline

\rule{0pt}{3ex}&\multicolumn{8}{c}{Fixed Effects with Robust SE}\\ \cline{2-9}
\rule{0pt}{3ex} &\multicolumn{2}{c}{\underline{Model 1}}&&\multicolumn{2}{c}{\underline{Model 2}}&&\multicolumn{2}{c}{\underline{Model 3}}\\
			&$\gamma$&\emph{p}&&$\gamma$&\emph{p}&&$\gamma$&\emph{p}\\

\multicolumn{3}{l}{ Intercept $\beta_{0j}$}	&&&&&&\\
~~Intercept $\gamma_{00}$		&55.05&\textless0.001	&&57.70&\textless0.001	&&54.52&\textless0.001	\\
~~LA-Sup. $\gamma_{01}$		&-&-					&&-&-					&&5.50&0.003	\\
~~Class Pre $\gamma_{02}$	&-&-					&&-&-					&&0.94&\textless0.001	\\
\multicolumn{9}{l}{For Underserved Gender slope $\beta_{1j}$} 	\\
~~Intercept $\gamma_{10}$	&-&-					&&-3.99&\textless0.001	&&-3.54&0.001	\\
~~LA-Sup. $\gamma_{11}$	&-&-					&&-&-					&&-0.25&0.862	\\
\multicolumn{9}{l}{For Underserved Race/Ethnicity slope $\beta_2$} 	\\
~~Intercept	$\gamma_{20}$	&-&-					&&-4.30&\textless0.001	&&-4.08&\textless0.001	\\
~~LA-Sup. $\gamma_{21}$	&-&-					&&-&-					&&-0.21&0.884	\\
\multicolumn{9}{l}{For Student Prescore slope $\beta_3$} 	\\
~~Intercept	$\gamma_{30}$	&-&-					&&0.52&\textless0.001	&&0.51&\textless0.001	\\\hline
\rule{0pt}{3ex}&\multicolumn{8}{c}{Random Effect Variance}\\ \cline{2-9}
\rule{0pt}{3ex} &\multicolumn{2}{c}{\underline{Model 1}}&&\multicolumn{2}{c}{\underline{Model 2}}&&\multicolumn{2}{c}{\underline{Model 3}}\\
Intercept $\textit{u}_{0j}$ &\multicolumn{2}{c}{163.99}&&\multicolumn{2}{c}{154.41}&&\multicolumn{2}{c}{69.85}\\
Gender  $\textit{u}_{1j}$&\multicolumn{2}{c}{-}&&\multicolumn{2}{c}{10.01}&&\multicolumn{2}{c}{3.26}\\
Race/Eth. $\textit{u}_{2j}$  &\multicolumn{2}{c}{-}&&\multicolumn{2}{c}{1.85}&&\multicolumn{2}{c}{1.36}\\
Prescore $\textit{u}_{3j}$ &\multicolumn{2}{c}{-}&&\multicolumn{2}{c}{0.01}&&\multicolumn{2}{c}{0.01}\\
Level-1 $r_{ij}$&\multicolumn{2}{c}{405.62}&&\multicolumn{2}{c}{297.42}&&\multicolumn{2}{c}{297.62}\\

\hline \hline

\end{tabular}
\end{table}





\section{Findings}

Model 3 reliably (p$\leq$0.001) identifies gaps in posttest scores across student demographics while controlling for student and class average pretest scores. In non-LA-supported courses, the model predicts that students from dominant and underserved genders who begin the class with the \textit{same} pretest scores will have a difference in posttest scores of 3.5\%. A similar gap (4.1\%) emerges between students from dominant and underserved races/ethnicities. The model predicts that students in non-LA-supported courses who are underserved by gender \textit{and} race/ethnicity will score 7.6\% lower than their peers in dominant gender and race/ethnicity groups with equivalent pretest scores.

\begin{figure}

\includegraphics[width=1\columnwidth]{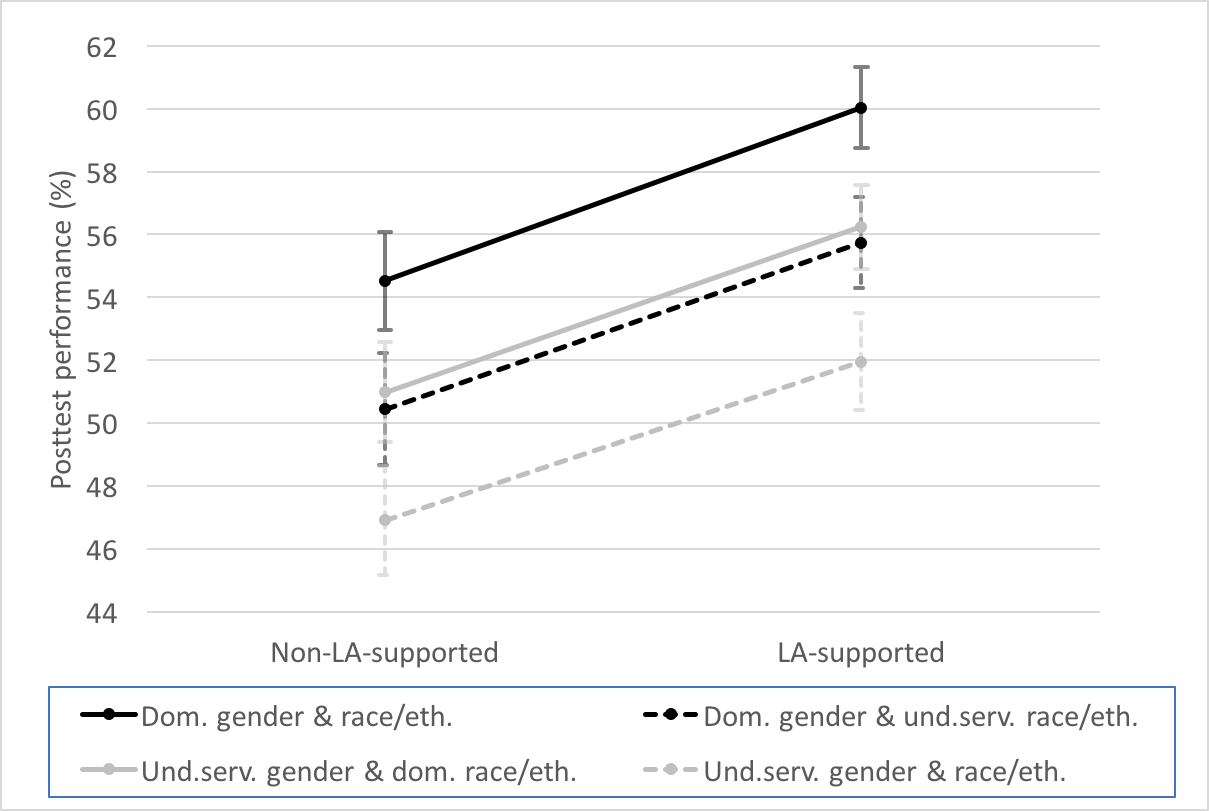}

\caption{Predicted posttest scores (+/- 1 S.E.) controlling for pretest scores}

\end{figure}

\par	Model 3 shows student posttest scores were higher in LA-supported courses than traditional courses across all demographics. While students performed better overall in LA-supported, the model shows the predicted performance gaps were not reliably (p\textgreater0.8) or meaningfully (\textit{d}$\sim$0.01) smaller than in non-LA-supported courses. Figure 1 shows the predicted posttest score for students with average pretest scores across demographic groups in non-LA and LA-supported courses. The differences in scores between groups of students are very similar in both settings.  While LAs were not associated with a reduction in the raw differences in average group posttest scores, they decreased the percentage in difference in student gains across groups. Courses with LAs have higher relative gains for students who are from underserved groups, 87\% versus 80\%, compared to the gain for students from dominant backgrounds.

\section{Discussion}

Differences in performance across genders in physics has been the focus of many investigations\citep{Brewe2016EditorialPhysics} and is well established. Research into the differences in performance across racial and ethnic lines in physics has received limited attention. Our results show that the inequities in performance by race and ethnicity are similar in size to the inequities by gender, indicating that these inequities deserve similar levels of attention. It is possible that the inequities by race and ethnicity have received little attention because underserved race/ethnicity students have so little representation in introductory physics courses that it has been very difficult for researchers to get reliable measures of these differences.

\par 	At first appearance the posttest difference for gender (3.5\%) and race/ethnicity (4.1\%) may seem like only a small difference. Given that the average improvement from pre to posttest for students with dominant in gender and race/ethnicity identities is approximately 20\%, falling behind by 3.5\% or 4.1\% over the course of a semester represents missing out on nearly a fifth of the average gain. Students who are underserved by gender \textit{and} race/ethnicity miss out on over one third of the gain of their peers from dominant groups. The population of students who are underserved by both gender and race/ethnicity is small in physics and thus it is very difficult to investigate their performance with quantitative methods. 

\par	Contrary to the findings in our exploratory investigation that did not utilize nested models\citep{White2016ThePhysics}, raw inequities in student posttest scores were effectively constant across non-LA and LA-supported contexts. Because LAs were associated with improved outcomes for all students, LAs reduced the relative gaps in gains between student groups. 

\section {Limitations and future work}

This investigation identified consistent, reliable, and meaningful inequities in student performance in introductory physics courses. While our findings showed no significant differences in the gaps within classroom contexts, it is unclear how representative our non-LA-supported course data is of introductory physics courses more broadly. The LASSO platform has been primarily promoted to faculty in the LA Alliance, which likely skewed the courses to be ones that use research-based pedagogical practices whether they were LA-supported or not. Thus, the inequities that are in the courses without LAs may not be representative of inequities in traditional lecture-based courses. We expect that increased adoption of the LASSO platform will improve the generalizability of our findings. In our future work we will use a meta-analysis of published results to further inform our analysis. This work is funded in part by NSF-IUSE Grant No. DUE-1525338 and is Contribution No. LAA-045 of the Learning Assistant Alliance.

 \bibliography{Mendeley}

\end{document}